\documentclass[oldversion,rnote]{aa} 
\usepackage{graphicx}
\usepackage{aalongtable}
\usepackage{txfonts}
\usepackage{rotating}
\usepackage{lscape}
\usepackage{natbib}
\newcommand{\gsim}{\;\lower.6ex\hbox{$\sim$}\kern-7.75pt\raise.65ex\hbox{$>$}\;}
\newcommand{\lsim}{\;\lower.6ex\hbox{$\sim$}\kern-7.75pt\raise.65ex\hbox{$<$}\;}

\begin{document}
\title{On the radial distribution of stars of different stellar generations in
the globular cluster NGC~3201\thanks{Based on observations collected at 
ESO telescopes under programme 073.D-0211}
 }

\author{
E. Carretta\inst{1},
A. Bragaglia\inst{1},
V. D'Orazi\inst{2},
S. Lucatello\inst{2,3,4},
\and
R.G. Gratton\inst{2}
}

\authorrunning{E. Carretta et al.}
\titlerunning{Radial distribution of stars in NGC~3201}

\offprints{E. Carretta, eugenio.carretta@oabo.inaf.it}

\institute{
INAF-Osservatorio Astronomico di Bologna, Via Ranzani 1, I-40127
 Bologna, Italy
\and
INAF-Osservatorio Astronomico di Padova, Vicolo dell'Osservatorio 5, I-35122
 Padova, Italy
\and
Excellence Cluster Universe, Technische Universit\"at M\"unchen, 
 Boltzmannstr. 2, D-85748, Garching, Germany 
\and
Max-Planck-Institut f\" ur Astrophysik, D-85741 Garching, Germany  }

\date{}

\abstract{We study the radial distribution of stars of different stellar
generations in the globular cluster NGC~3201. 
From recently published 
multicolour photometry, a radial dependence of the location of
stars on the giant branch was found. We coupled the photometric information
to  our sample of 100
red giants with Na, O abundances and known classification as first or
second-generation stars.
We find that giants stars of the second generation in
NGC~3201 show a tendency to be more centrally concentrated than stars of the
first generation, supporting less robust results from our spectroscopic
analysis.
  }

\maketitle

\section{Introduction}

Chemical inhomogeneities in light elements (C, N, O, F, Na, Al, Mg, and even Si)
are intrinsic to globular clusters (GCs; see \citealt{araa04} for an extensive
review and references). In particular, the striking anticorrelation between Na
and O abundances in GC red giant branch (RGB) stars discovered and studied by
the Texas-Lick group (see \citealt{kraft94} for a review on those pioneering efforts)
is the most notable signature observed with high resolution  spectroscopy. The
pivotal discovery of the Na-O (and the Mg-Al)  anticorrelation in unevolved
cluster stars \citep{gratton01} led to the unambiguous conclusion that stars of
different generations  co-exist in the currently observed GCs. The reason is
that the high temperatures  required for proton-capture reactions
\citep{denden89,langer93}  to produce matter depleted in O, Mg and enriched in
Na, Al are never reached in the interior of low-mass stars (temperature in
excess of 25 or 70$\times10^6$ K are required for the NeNa and MgAl cycles,
respectively).
As a consequence, the observations by Gratton et al.  \citep[later confirmed by
e.g.,][]{cohen02,carretta04,dorazi10} require the existence of a previous
stellar generation of more massive stars with higher
internal temperatures sufficient to activate the necessary
nucleosynthesis, providing the ejecta from which the
second-generation stars formed. Dilution processes with pristine gas left in the
cluster may then reproduce the whole length of the Na-O anticorrelation
(see \citealt{pc06}; see however \citealt{gra010}).

The Na-O anticorrelation is so widespread among GCs \citep[see e.g.,][]{carretta09a,carretta09b}
that this feature is
probably  associated to the very same mechanism of cluster
formation \citep{carretta06} and may be considered the main criterion to
discriminate between {\it bona fide} GCs \citep{carretta10} and other type
of clusters, regardless of their old ages or even large mass 
\citep[e.g. NGC~6791,][in prep.]{bragaglia6791}.

However, while the overall pattern of the chemical composition of first and
second-generation GC stars is currently well assessed, several issues are still
left open, the principal being the nature of the actual polluters,
either intermediate-mass Asymptotic Giant Branch stars 
\citep[IM-AGBs,][]{ventura01}  or
fast-rotating massive stars \citep[FRMSs,][]{decressin07}.

One of the main questions concerns the possible link between chemical signature
and dynamical evolution of different stellar generations in GCs. This issue is
puzzling and still poorly explored in a systematic way by theoretical models.
Models where a cooling-flow feeds gas enriched in IM-AGB ejecta to
form second generation stars \citep{dercole08}, intrinsically predict that
this generation should be more centrally  concentrated, since the gas is
re-collected at the cluster centre.  On the other hand, also second-generation
stars formed by matter polluted by FRMS are expected to be more centrally
concentrated. They should have the same radial distribution of their
progenitors, that are assumed to be born near the cluster centre, being very
massive objects. 

Even if the second generation formed at the cluster centre, there is the 
action of the dynamical evolution over a Hubble time to be taken into account.
For instance, \cite{decressin08} compute that the second-generation stars 
are progressively spread out by dynamical encounters. As a result, the radial 
distributions of first and second-generation stars can no longer be
distinguished from their dynamics alone at the present time. 
In this Note we combine information from our ongoing FLAMES survey  of chemical
abundances in giants in GCs \cite[see][]{carretta06} with newly derived
photometry for the GC \object{NGC 3201} 
\citep{kravtsov09,kravtsov10} and we  provide new insights on the radial
distribution of first and second-generation stars in this GC, whose relaxation
time at half-mass radius is about 1.6 Gyr \cite{har96}.

\section{Radial distribution of P, I, and E stars in NGC~3201}

In \cite{carretta09a} we showed that  it is possible to
separate a stellar component of first-generation (or primordial, P) stars
in all GCs observed. The
remaining second-generation stars can be further separated into intermediate (I)
and extreme (E) components, according to the degree of O-depletion and
Na-enhancement along the Na-O anticorrelation. 

We also tried to study the radial distribution of P, I, and E stars from the
total sample of almost 2000 RGB stars in 19 GCs, but  concluded that the
merged sample is not the best option. Owing to the physical limitations of the
FPOSS positioner of the FLAMES fibres, and to different mass, concentration,
and size of the GCs, we were not able to sample cluster regions that
are dynamically equivalent in all objects. As a consequence, the cumulative
distributions from the whole sample in \cite{carretta09a} may be biased.

This bias is partly reduced when looking at an individual cluster. As an example we select
NGC~3201, since it hosts a good fraction of each component among the 100 stars
with Na, O abundances (P=$35\pm6\%$, I=$56\pm7\%$, and E=$9\pm3\%$,
\citep{carretta09a}). This
cluster is also the subject of a recent publication advocating inhomogeneity of
its stellar population on photometric basis
\citep[][see below]{kravtsov10}.

\begin{figure}
\centering
\includegraphics[scale=0.40]{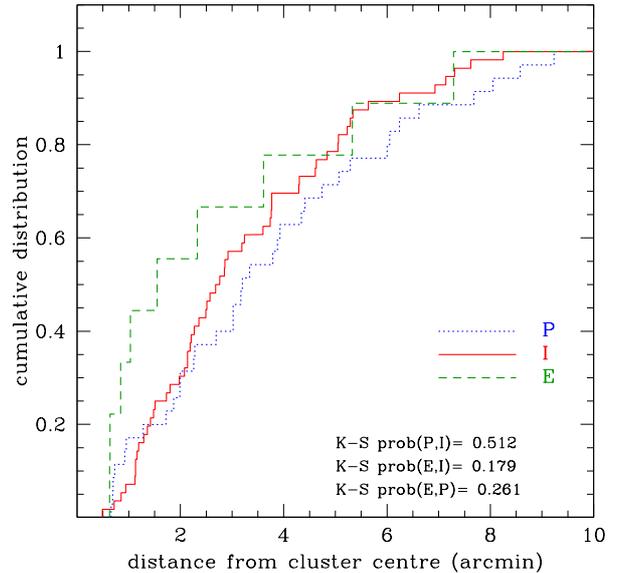}
\caption{Cumulative distributions of the primordial P (blue dotted line), the
intermediate I (red solid line), and the extreme E stellar components (green 
dashed line) in NGC~3201 as a function of the distance from the cluster centre.
The results of the comparison between the distributions using a K-S test are
also listed.}
\label{f:distrPIE}
\end{figure}

\begin{figure}
\centering
\includegraphics[scale=0.40]{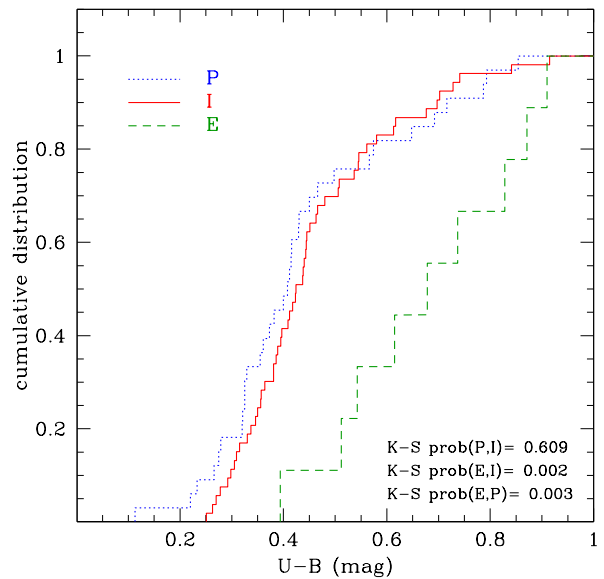}
\caption{Cumulative distributions of $U-B$ colours of P (blue dotted line), I
(red solid line), and E (green dashed line) stars in NGC~3201.
The results of the comparison between the distributions using a K-S test are
also listed.}
\label{f:distrPIEcol}
\end{figure}

In Fig.~\ref{f:distrPIE} we show the cumulative distributions of radial
distances for stars of the P, I, and E components in NGC~3201.
Although it is possible to see a trend for second-generation stars to be more
centrally concentrated (and progressively more the E than the I component),  
the distributions show no statistically
significant differences if we apply a Kolmogorov-Smirnov (K-S) test.

Much more significant are the differences in colour observed along the RGB for
the P, I, and E populations, in particular when using the $U-B$ colour. 
We do not have $U$ in our photometric database, 
so we cross-identified our stars with the multi-colour photometry of more 
than 12000 stars in NGC~3201 by \cite{kravtsov09}. We 
used a software written by P. Montegriffo at the Bologna Observatory and
found 138 stars (out of 150) in 
common between the two samples. Among these, 95 objects\footnote{ We required
both Na and O values for the classification, so stars for which one of the two
is not available were not considered. However, a less strict classification in
first and second-generation (without distinction between I and E) is possible
whenever Na is measured \citep[see e.g.][]{bragaglia10}.} can be classified as
P, I, or E RGB stars, and their distributions in $U-B$ colour are shown in
Fig.~\ref{f:distrPIEcol}.

\begin{figure*}
\centering
\includegraphics[bb=20 146 588 572, scale=0.60]{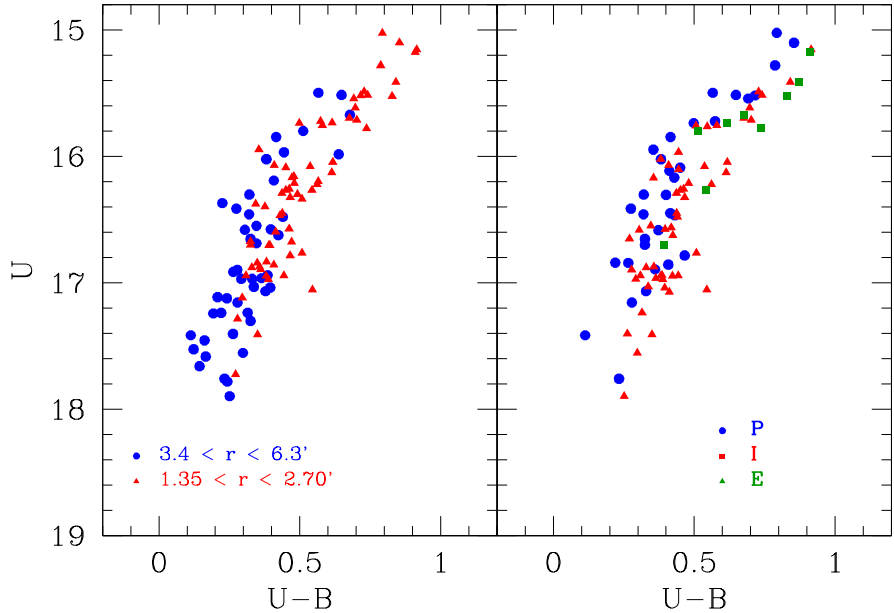}
\caption{Left panel: $U,U-B$ CMD \citep{kravtsov09} for RGB stars of  NGC~3201
in common with \cite{carretta09a,carretta09b}. Blue circles and red  filled
triangles  corresponds to the two different intervals in radial distances used
in \cite{kravtsov10}. Right panel: the same CMD using stars in common, but with
symbols are colour-coded  according to the P (blue circles), I (red triangles),
and E (green squares)  components of  first and second-generation stars.}
\label{f:rgbPIE}
\end{figure*}

In this case, while the P and I distributions are not statistically different,
there is no doubt that the E stars have colours redder than the other two
components, at a very high level of confidence. This is not a new result: 
\cite{marino08}, \cite{yong08}, and \cite{carretta09a}  
convincingly showed that in M~4 and NGC~6752 Na-poor/O-rich/N-poor first-generation stars are
segregated on the RGB into a narrow strip to the blue ridge of the branch,
either in the $U,U-B$ colour-magnitude diagram (CMD) 
or in the $u,u-b$ CMD from Str\"omgren photometry.
Second-generation stars (I+E, Na-rich/O-poor, and N-rich) are instead spread out
to the red of the RGB.  This different location is due to the effect of N
abundances, affecting $U-B$ and $u-b$ colour indexes through the formation of NH
and CN molecular bands falling into the windows of the ultraviolet 
filters response.

Fig.~\ref{f:distrPIEcol} simply confirms this behaviour for yet another GC, by
coupling colours and Na/O abundances. However, the novelty presented in this
Note is that we can exploit this property to get an independent insight on the
radial distributions of stars in NGC~3201 which is completely free from
any bias in the spectroscopic sample.

\cite{kravtsov10} found signs of multiple sequences among subgiant branch stars in
this GC, with stars at different magnitude levels preferentially lying at
different radial distances, with a high level of statistical significance.
Additionally, a similar dependence on the radial distances is observed among RGB
stars, with the bluer stars in $U-B$ systematically located towards outer
cluster regions \citep[see][their Fig.3]{kravtsov10}.

We show in Fig.~\ref{f:rgbPIE} (left panel) the $U,U-B$ CMD for all
the 138 stars of our
sample  in NGC~3201 in common with Kratsov et al. The colour coding identifies
the same two intervals in radial distances used by them and reflects the trend
noted in the radial distribution. In the right panel, however, the stars in
common\footnote{Note that at the faintest magnitudes some stars are missing in
the right panel, in particular on the blue side. This happens because of the
increased  difficulty to measure reliable equivalent widths for faint and warm
stars. In \cite{carretta09a} only stars with both Na and O abundances are
attributed to one of the P, I, E populations.} are simply colour-coded according
to their status of P, I, or E. The comparison between the two plots immediately
tells us that second-generations stars in NGC~3201 must also be those more
concentrated toward the cluster centre, according to their location on the CMD. 

This result supports  what we find from our sample only (see
Fig.~\ref{f:distrPIE}), but is now more robust, being anchored to a much
larger sample (almost 500 stars with $UBVI$ photometry with respect to about
100 RGB stars with Na,O abundances).

In conclusion, we find that  second-generation stars are most likely
more concentrated toward the inner cluster regions in NGC~3201. Apparently, 
it seems that
even if the relaxation time of NGC~3201 is much shorter than its lifetime, some
trace of the initial formation process might be still recognizable. 
Interestingly, \cite{norris79} found evidence of the same effect in another cluster, 
47 Tucane: they determined that the CN-strong stars, i.e., the ones that we now attribute 
to the second generation, are more centrally located.

While our finding does not
help to discriminate between scenarios where the pollution comes either from AGB
or FRMS, it offers another piece of evidence, useful to reconstruct the
complex puzzle of multiple stellar populations in GCs.

\begin{acknowledgements}
This work was partially funded by the Italian MIUR under PRIN 2003, 2007, and
by the grant INAF 2005 ''Experimenting
nucleosynthesis in clean environments''. 
This research has made use of the SIMBAD database (in particular 
Vizier), operated at CDS,
Strasbourg, France and of NASA's Astrophysical Data System.
\end{acknowledgements}


\begin{thebibliography}{}

\bibitem[Bragaglia et al.(2010a)]{bragaglia10}
Bragaglia, A., Carretta, E., Gratton, R., D'Orazi, V. Cassisi, S., \& Lucatello,
S. 2010a, \aap, in press

\bibitem[Bragaglia et al.(2010b)]{bragaglia6791} 
Bragaglia et al. 2010b, in preparation 

\bibitem[Carretta(2006)]{carretta06} 
Carretta, E. 2006, \aj, 131, 1766 

\bibitem[Carretta et al.(2006)]{paperI} 
Carretta, E., Bragaglia, A., Gratton R.G., Leone, F., 
Recio-Blanco, A., \& Lucatello, S. 2006, \aap, 450, 523 

\bibitem[Carretta et al.(2009a)]{carretta09a}  
Carretta, E. et al. 2009a, \aap, 505, 117  

\bibitem[Carretta et al.(2009b)]{carretta09b} 
Carretta, E., Bragaglia, A., Gratton, R.G., \& Lucatello, S. 2009b, 
 \aap, 505, 139   

\bibitem[Carretta et al.(2010)]{carretta10} 
Carretta, E., Bragaglia, A., Gratton, R.G., Recio-Blanco, A.,
Lucatello, S., D'Orazi, V., \& Cassisi, S. 2010a, \aap, in press, arXiv:1003.1723 

\bibitem[Carretta et al.(2004)]{carretta04} 
Carretta, E., Gratton, R.~G., Bragaglia, A., Bonifacio, P., \& Pasquini, L.\
2004, \aap, 416, 925


\bibitem[Cohen et al.(2002)]{cohen02} 
Cohen, J.~G., Briley, 
M.~M., \& Stetson, P.~B.\ 2002, \aj, 123, 2525 

\bibitem[Decressin et al.(2008)]{decressin08} 
Decressin, T., Baumgardt, H., \& Kroupa, P. 2008, \aap, 492, 101

\bibitem[Decressin et al.(2007)]{decressin07} 
Decressin, T., Meynet, G., Charbonnel C. Prantzos, N.,\& 
Ekstrom, S. 2007, \aap, 464, 1029 

\bibitem[Denisenkov \& Denisenkova(1989)]{denden89} 
Denisenkov, P.A.,\&  Denisenkova, S.N. 1989, A.Tsir., 1538, 11

\bibitem[D'Ercole et al.(2008)]{dercole08} 
D'Ercole, A., Vesperini, E., D'Antona, F., McMillan, S.L.W., \& 
 Recchi, S. 2008, MNRAS, 391, 825

\bibitem[D'Orazi et al.(2010)]{dorazi10} 
D'Orazi, V., Lucatello,  S., Gratton, R., Bragaglia, A., Carretta, E., Shen,
Z.,  \& Zaggia, S.\ 2010, \apjl, 713, L1   

\bibitem[Gratton \& Carretta(2010)]{gra010} Gratton, R.G., Carretta, E. 2010,
submitted to A\&A

\bibitem[Gratton et al.(2001)]{gratton01} 
Gratton, R.G. et al. 2001, \aap, 369, 87

\bibitem[Gratton et al.(2004)]{araa04} 
Gratton, R.G., Sneden, C., \& Carretta, E. 2004, \araa, 42, 385

\bibitem[Harris (1996)]{har96} Harris, W.~E. 1996, AJ, 112, 1487

\bibitem[Kraft(1994)]{kraft94} 
Kraft, R.~P. 1994, PASP, 106, 553 

\bibitem[Kravtsov et al.(2009)]{kravtsov09} 
Kravtsov, V., Alcaino, G., Marconi, G.,\&  Alvarado, F. 2009, \aap, 497, 371 

\bibitem[Kravtsov et al.(2010)]{kravtsov10} 
Kravtsov, V., Alcaino, G., Marconi, G., \& Alvarado, F. 2010, \aap, 512, L6 

\bibitem[Langer et al.(1993)]{langer93} 
Langer, G.E., Hoffman, R., \& Sneden, C. 1993, \pasp, 105, 301

\bibitem[Marino et al.(2008)]{marino08}
 Marino, A.Villanova, S., Piotto, G., Milone, A.P., Momany, Y., Bedin, L.R., 
 \& Medling, A.M. 2008, \aap, 490, 625

\bibitem[Norris \& Freeman(1979)]{norris79} Norris, J., \& Freeman, K.~C.\ 1979, \apjl, 230, L179 

\bibitem[Pranztos \& Charbonnel(2006)]{pc06} 
Prantzos, N.,\&  Charbonnel, C. 2006, \aap, 458, 135 

\bibitem[Ventura et al.(2001)]{ventura01} 
Ventura, P. D'Antona, F., Mazzitelli, I., \& Gratton, R. 2001,
  \apj, 550, L65 

\bibitem[Yong et al.(2008)]{yong08} 
Yong, D., Grundahl, F., Johnson, J.A.,\&  Asplund, M. 2008, \apj, 684,
  1159 

\end{thebibliography}
\end{document}